\documentclass[twocolumn]{aastex7}
\usepackage{hyperref}
\hypersetup{colorlinks, citecolor=blue, filecolor=blue, linkcolor=blue, urlcolor=blue}
\usepackage{amsmath}
\usepackage{longtable}
\usepackage{comment}
\usepackage{graphicx}
\usepackage{subfigure}
\usepackage{textcomp}
\usepackage{longtable}
\usepackage{soul}
\usepackage{xcolor}
\sethlcolor{yellow}
\usepackage[most]{tcolorbox}
\usepackage{threeparttable}
\usepackage{tabularx}
\usepackage{placeins}

\begin{document}

\title{CW Cas: A solar-type contact binary system with an unseen third companion in a hierarchical quadruple system}

\correspondingauthor{Zhu LiYing}
%\author{Author\thanks{Corresponding author.}}
\email[show]{zhuly@ynao.ac.cn}

\author[orcid=0000-0003-2999-6879,sname='Matekov']{Azizbek Matekov}
\affiliation{Yunnan Observatories, Chinese Academy of Sciences, Kunming 650216, People's Republic of China}
\affiliation{University of Chinese Academy of Sciences, No. 1 Yanqihu East Road, Huairou District, Beijing 101408, People's Republic of China}
\affiliation{Ulugh Beg Astronomical Institute, Uzbekistan Academy of Sciences, 33 Astronomicheskaya Street, Tashkent 100052, Uzbekistan}
\email[show]{azizbek.matekov@ynao.ac.cn}  

\author[orcid=0000-0002-0796-7009,gname=Liying, sname='Zhu']{Liying Zhu} 
\affiliation{Yunnan Observatories, Chinese Academy of Sciences, Kunming 650216, People's Republic of China}
\affiliation{University of Chinese Academy of Sciences, No. 1 Yanqihu East Road, Huairou District, Beijing 101408, People's Republic of China}
\affiliation{Key Laboratory of the Structure and Evolution of Celestial Objects, Chinese Academy of Sciences, Kunming 650216, People's Republic of China}
\email{zhuly@ynao.ac.cn}

\author[orcid=0000-0003-3767-6939, gname=Li, sname='Linjia']{Linjia Li}
\affiliation{Yunnan Observatories, Chinese Academy of Sciences, Kunming 650216, People's Republic of China}
\affiliation{Key Laboratory of the Structure and Evolution of Celestial Objects, Chinese Academy of Sciences, Kunming 650216, People's Republic of China}
\email{lipk@ynao.ac.cn}

\author[orcid=0000-0002-8320-8469, gname=Meng,sname='Fangbin']{Fangbin Meng}
\affiliation{Yunnan Observatories, Chinese Academy of Sciences, Kunming 650216, People's Republic of China}
\affiliation{University of Chinese Academy of Sciences, No. 1 Yanqihu East Road, Huairou District, Beijing 101408, People's Republic of China}
\affiliation{Key Laboratory of the Structure and Evolution of Celestial Objects, Chinese Academy of Sciences, Kunming 650216, People's Republic of China}
\email{mengfangbin@ynao.ac.cn}

\author[orcid=0000-0002-2276-6352,sname=Nianping,gname='Liu']{Nianping Liu}
\affiliation{Yunnan Observatories, Chinese Academy of Sciences, Kunming 650216, People's Republic of China}
\affiliation{Key Laboratory of the Structure and Evolution of Celestial Objects, Chinese Academy of Sciences, Kunming 650216, People's Republic of China}
%\affiliation{Center for Astronomical Mega-Science, Chinese Academy of Sciences, 20A Datun Road, Chaoyang District, Beijing, 100012, PR China}
\email{lnp@ynao.ac.cn}

\author[orcid=0009-0009-7581-9209,sname=Qian,gname='ShengBang']{ShengBang Qian}
\affiliation{Department of Astronomy, School of Physics and Astronomy, Yunnan University, Kunming 650091, People's Republic of China}
\affiliation{Key Laboratory of Astroparticle Physics of Yunnan Province, Yunnan University, Kunming 650091, People's Republic of China}
\email{qiansb@ynu.edu.cn}

\author[orcid=0009-0009-7581-9209,sname=Zhao,gname='Ergang']{Ergang Zhao}
\affiliation{Yunnan Observatories, Chinese Academy of Sciences, Kunming 650216, People's Republic of China}
\affiliation{University of Chinese Academy of Sciences, No. 1 Yanqihu East Road, Huairou District, Beijing 101408, People's Republic of China}
%\affiliation{Center for Astronomical Mega-Science, Chinese Academy of Sciences, 20A Datun Road, Chaoyang District, Beijing, 100012, PR China}
\email{zergang@ynao.ac.cn}

\author[orcid=0000-0002-9975-7833,sname=Kovalev,gname='Mikhail']{Mikhail Kovalev}
\affiliation{Yunnan Observatories, Chinese Academy of Sciences, Kunming 650216, People's Republic of China}
\affiliation{Key Laboratory of the Structure and Evolution of Celestial Objects, Chinese Academy of Sciences, Kunming 650216, People's Republic of China}
\affiliation{International Centre of Supernovae, Yunnan Key Laboratory, Kunming 650216, China}
\email{mikhail.kovalev@ynao.ac.cn}

\author[sname=Jiangjiao,gname='Wang']{Jiangjiao Wang}
\affiliation{Yunnan Observatories, Chinese Academy of Sciences, Kunming 650216, People's Republic of China}
\affiliation{University of Chinese Academy of Sciences, No. 1 Yanqihu East Road, Huairou District, Beijing 101408, People's Republic of China}
%\affiliation{Center for Astronomical Mega-Science, Chinese Academy of Sciences, 20A Datun Road, Chaoyang District, Beijing, 100012, PR China}
\email{wangjiangjiao@ynao.ac.cn}

\author[sname=Yufei,gname='Chen']{Yufei Chen}
\affiliation{Yunnan Observatories, Chinese Academy of Sciences, Kunming 650216, People's Republic of China}
\affiliation{University of Chinese Academy of Sciences, No. 1 Yanqihu East Road, Huairou District, Beijing 101408, People's Republic of China}
\email{chenyufei@ynao.ac.cn}

\author[orcid=0009-0008-8819-7640]{Aktam Khafizov}
\affiliation{Ulugh Beg Astronomical Institute, Uzbekistan Academy of Sciences, 33 Astronomicheskaya Street, Tashkent 100052, Uzbekistan}
\email{aktam@astrin.uz}

\author[orcid=0000-0001-5511-7183]{Soonthornthum Boonrucksar}
\affiliation{National Astronomical Research Institute of Thailand, 191 Siriphanich Building., Huay Kaew Road, Chiang Mai 50200, Thailand}
\email{boonrucksar@narit.or.th}

\author{Somsawat Rattanasoon}
\affiliation{National Astronomical Research Institute of Thailand, 191 Siriphanich Building., Huay Kaew Road, Chiang Mai 50200, Thailand}
\email{somsawat@narit.or.th}

\author[orcid=0000-0001-9730-3769]{Shuhrat Ehgamberdiev}
\affiliation{Ulugh Beg Astronomical Institute, Uzbekistan Academy of Sciences, 33 Astronomicheskaya Street, Tashkent 100052, Uzbekistan}
\affiliation{National University of Uzbekistan, 4 University Street, Tashkent 100174, Uzbekistan}
\email{shuhrat@astrin.uz}

%\collaboration{all}{The Terra Mater collaboration}

\begin{abstract}
We present a comprehensive multiband photometric and spectroscopic study of the G-type binary CW Cas. Analysis of its double-lined spectroscopic radial velocity curves yields a reliable mass ratio of $q = 1.88(9)$. By combining modeling of the radial velocity curves with $BVR_{c}I_{c}$ and TESS light curves, we found that CW Cas is a W-subtype shallow contact binary with a fill-out factor of 15\%. The components have masses of $0.98(6)M_{\odot}$ and $0.52(4)M_{\odot}$, separated by $2.25(5)R_{\odot}$. A notable asymmetry in the maxima of the light curves was detected and explained by a dark spot located on the surface of at least one component. Comparison of light curves from different years revealed that these dark spot activities exhibit cyclic variations with an approximate period of 349 days. Orbital period analysis via its \mbox{\it O\,--\,C} diagram spanning 125 yr shows a long-term decrease superimposed with a periodic oscillation caused by the light-travel time effect due to a third body. This tertiary component has an orbital period of $P_{3} = 99.4(6)$ yr and a minimal mass of $M_{3}=0.91(1)M_{\odot}$. The absence of detectable signatures for this massive object in either the spectroscopic or photometric data sets implies it should be a compact object. Furthermore, a visual companion was identified based on Gaia Data Release 3 astrometric data, suggesting that CW Cas is part of a hierarchical quadruple system. As such, CW Cas represents a valuable laboratory for probing a 2+1+1 hierarchical multiple system hosting a compact object.
\end{abstract}



\section{Introduction} \label{sec1}

W UMa-type contact binary systems usually contain two late-type main-sequence stars that fill their critical Roche lobes \citep{Kopal_1955AnAp...18..379K}, and share a common convective envelope (CCE) \citep{Lucy_1968ApJ...151.1123L}, creating a unique and complex astronomical phenomenon \citep{Yildirim_2025NewA..12102445Y}.
%Their light curves (LCs) have short periods ($p \lesssim 0.5$ days), vary smoothly, and are characterized by wide eclipses. The eclipse depths are also very similar, resulting in an inferred temperature ratio close to unity, independent of the system's mass ratio. Due to the proximity of the components, the inclination of the contact binaries can be relatively low while still producing partial eclipses \citep{FabryPrsa_2025ApJ...994....7F}. 
The light curve of this contact binary displays a typical EW-type variation, characterized by nearly equal-depth primary and secondary eclipses and continuous brightness variations. This morphology suggests that the two components have similar surface temperatures (i.e., a temperature ratio close to unity) and exhibit distorted shapes due to their close interaction. Furthermore, the proximity of the stars enables partial eclipses even at relatively low orbital inclinations. \citep{Qian_2017RAA....17...87Q, FabryPrsa_2025ApJ...994....7F}. \citet{Binnendijk_1970VA.....12..217B} classified W UMa-type contact binaries into two subtypes i.e. A-subtype and W-subtype, based on their light curve characteristics. The key distinction lies in the relative surface temperatures of the components: in A-subtype systems, the more massive star is hotter, whereas in W-subtype systems, the less massive component exhibits a higher temperature.

Statistical analyses of contact binaries \citep{Eggen_1967MmRAS..70..111E, Rucinski_1998AJ....116.2998R, Qian_2020RAA....20..163Q, Pribulla_2006AJ....132..769P}, reveal that W UMa systems typically have short orbital periods (less than one day). Among these, G-type systems are the most numerous subgroup, and a significant fraction show evidence of a third body in the system. As these binaries typically consist of two late-type stars, their light curves often display unequal heights of the primary and secondary maximum—a phenomenon known as the O'Connell effect \citep{OConnell1951MNRAS.111..642O}. The most widely accepted explanation for this asymmetry involves magnetic activity, such as starspots on one or both components. Long-baseline, continuous photometric data from ground-based surveys like the All-Sky Automated Survey for Supernovae (ASAS-SN; \citealt{Shappee_2014ApJ...788...48S, Christy_2023MNRAS.519.5271C}) and space-based missions including the Transiting Exoplanet Survey Satellite (TESS; \citealt{Ricker_2015JATIS...1a4003R}) and Kepler (\citealt{Borucki_2010Sci...327..977B})—provide exceptional opportunities to investigate the temporal evolution of the O'Connell effect over extended timescales. However, these asymmetries introduce significant uncertainties when deriving physical parameters from light curve modeling alone. Therefore, combining photometric observations with radial velocity (RV) measurements is essential for these active binaries to obtain accurate stellar parameters, and benefit to precisely characterize potential third bodies as well.

%All of these factors make contact binary systems fundamental astrophysical laboratories for studying stellar structure, formation mechanisms, and evolutionary pathways \citep{Yildirim_2025NewA..12102445Y, WangYiFan_2025MNRAS.544.1958W}. }  

%The close distance between system components of the EW-type binaries light curves vary smoothly, and in the ideal case both maximum light levels are equal. In late-type stars, relatively strong magnetic activity is observed; the light curves of contact binaries usually show unequal heights of the primary and secondary maximum, i.e. the O'Connell effect \citep{OConnell1951MNRAS.111..642O}.  
%Most of the EW-type binaries orbital period shorter than 1 day and period distribution peak around 0.3 days \citep{Qian_2017RAA....17...87Q}. 
  %Especially based on long time span and continuous photometric data from sky survey such as ASAS-SN \citep{Shappee_2014ApJ...788...48S, Christy_2023MNRAS.519.5271C} and space telescopes such as Kepler \citep{Borucki_2010Sci...327..977B} and TESS \citep{Ricker_2015JATIS...1a4003R}, in addition to the O'Connell effect, other asymmetries, such as the shape, timing, and depth of the minima, have been detected in light curves of close binaries. 

CW Cassiopeia (CW Cas, TIC 421110675) was discovered as a variable star by \cite{Zverev1938}, with a period of 0.31886304 days.
%(see other designations and basic info in Table \ref{table1}). 
The light curves of CW Cas were obtained from examined photoelectric data from 1957 to 1961, \citet{Broglia_1964MmSAI..35...23B} identified that the system is an EW-type contact binary. Based on photometric data and the color intrinsic method, it was determined that the spectral type of CW Cas is G8 \citep{Hilditch_1975MmRAS..79..101H, Hill_1975MmRAS..79..131H}. Using different methods, \citet{Burchi_1977ApSS..47...35B} analyzed the photoelectric observations of CW Cas in the Johnson V band originally conducted by \citep{Burchi_1975IBVS..962....1B}; however, they did not obtain consistent results. \citet{Barone_1988AA...197..347B} improved the method and analyzed the same dataset used by \citet{Burchi_1977ApSS..47...35B}, ultimately deriving a W-subtype solution with a mass ratio $q=1.85$. The light curve examined in \cite{Davidge_1984ApJS...55..571D} and reported by \cite{Burchi_1975IBVS..962....1B} determined that CW Cas exhibits the O'Connell effect. \citet{Pribulla_2001CoSka..31...26P} analyzed its light curves in the $BV$-band and determined that the fill-out degree was $f=2.2\%$ and the temperature difference $\Delta T = 424$ K between the components, indicating poor thermal contact between components. \cite{Jiang_2010PASJ...62..457J} analyzed light curves of CW Cas based on the CCD photometric data in November 2000 and concluded that CW Cas belongs to W-subtype contact binary with a mass ratio $q=2.234$, and a fill-out degree of $f=6.5\%$. From \mbox{\it O\,--\,C} curve analysis, they found a period decrease at a rate of $dP/dt =-3.44 \times 10^{-8}$ d yr$^{-1}$, and a cyclic variation with period of 63.7 years which was possibly caused by the light-time effect of a third body. \citet{Wang_2014AJ....148...95W} presented multiband light-curve observations of CW Cas and analyzed the V-band data obtained in 2004 and 2011 to determine the presence of spot activity. Their re-investigation of the \mbox{\it O\,--\,C} curves for all minima showed that the system's period exhibits a cyclic variation with a period of $P_{3} = 69.9$ years, superimposed on a linear increase. They attributed the cyclic variation to the light-travel time effect (LTTE), indicating the presence of a tertiary companion. The investigation by \citep{Stepien_2001AA...370..157S, LiuXray_2022AA...663A.115L} showed that CW Cas is an X-ray source, which indicates that this binary has strong dynamo-related activity. We have compiled general information about CW Cas from the literature and present it in Table \ref{table1}.

%\begin{comment}
\begin{table}[htb!]
	\centering
	\caption{Main literature data for CW Cas.}
	\begin{tabular*}{\columnwidth}{@{\extracolsep{\fill}}lcc}
		\hline\hline
		Parameter & Value & Source \\
		\hline
        \textbf{Identifying information} &  &  \\
        %GCVS & V*CW Cas & 8 \\
        Tycho-2 catalogue & TYC 4020-1743-1 & 6 \\
        ASASSN-V & J004552.48+630508.5 & 1 \\
        TESS input catalogue & TIC421110675 & 1 \\
        %AAVSO UID & 000-BDJ-618 & 1 \\
		Gaia DR3 source id & 523846239000492928 & 2 \\
		RA (J2000) & 00:45:52.704 & 2 \\
		Dec (J2000) & +63:05:08.456 & 2 \\
		$\mu_{\alpha}$ (mas yr$^{-1}$) & -83.567(1) & 2 \\
		$\mu_{\delta}$ (mas yr$^{-1}$) & 3.053(1) & 2 \\
		$\varpi$ (mas) & 6.0112(161) & 2 \\
		Distance (pc) & $165.27_{164.74}^{165.77}$ & 4  \\
		RUWE & 1.073 & 2, 7 \\
		$E(B-V)$ & 1.425 & 5 \\
        $B-V$ & 0.833 & 8 \\
        $T_{\rm{eff}}$ (K) & 5078.36 & 3 \\
        $T_{\rm{eff}}$ (K) & 5197.0(248.5) & 1 \\
                              &               &   \\
        \textbf{Photometric properties}  &    &  \\
        $TESS$ (mag)  & 10.65(3)  & 1 \\
        $Gaia$ (mag)  & 11.161(9)  & 1 \\
        $G_{BP}$ (mag)  & 11.551(31)  & 1 \\
        $G_{RP}$ (mag)  & 10.514(29)  & 1 \\
        $B_{T}$ (mag)  & 12.286(132)  & 1 \\
        $V_{T}$ (mag)  & 11.401(78)  & 1 \\
        $J$ (mag)  & 9.496(23)  & 1 \\
        $H$ (mag)  & 9.077(28)  & 1 \\
        $K$ (mag)  & 8.937(23)  & 1 \\
        
		\hline
	\end{tabular*}
	\label{table1}
    \begin{tablenotes}
			\item \rm{Sources: (1) \citep{Stassun_2018AJ....156..102S}, (2)  \citep{Gaia_2022yCat.1355....0G}, (3) \citep{Gaia_2018yCat.1345....0G}, (4) geometric distance from \citep{BailerJones_2021AJ....161..147B}, (5) \citep{Shappee_2014ApJ...788...48S, Christy_2023MNRAS.519.5271C}, (6) \citep{HogTYCHO_2000AA...355L..27H}, (7) The Gaia Renormalized unit weight error (RUWE) is the square root of the normalized $\chi^2$ of the astrometric fit to the along-scan observations. Values in excess of about unity are some
            times taken to be a sign of stellar multiplicity. (8) \citep{Terrel_2012AJ....143...99T} }
	\end{tablenotes}
\end{table}
%\end{comment}

%Photometric data for an extended period are crucial for studying the orbital dynamics of binary systems. At the same time, high-precision data can reveal new asymmetries, and these asymmetries over time introduce additional uncertainty in deriving physical parameters from light curve analysis. In this context, further research combined with radial velocity studies is essential to obtain reliable physical parameters of the contact binary system and to detect a potential third body within the system.

%\textbf{In this article we present a multiband photometric and first spectroscopic analysis of the CW Cas. In Sec. \ref{section2} discuss observations of new multi-band observations, the first radial velocity measurements, and TESS light curves, and assume data collections. Additionally, we analyze the \mbox{\it O\,--\,C} diagram of the brightness minima, incorporating new eclipse timing data spanning over 125 years in Sec. \ref{section3}. We investigate the orbital and atmospheric parameters, as well as the periodic activity of starspots (O'Connell effect) on the stellar surface in Sec. \ref{section4}. We also explore the potential presence of an unseen massive third body—undetectable in spectral observations but suggested by the \mbox{\it O\,--\,C} diagram analysis. Furthermore, we examine the likelihood that the binary system is part of a quadruple system, based on Gaia data and our spectral observations in Sec. \ref{section5}.}

In this paper, we present the first radial velocity (RV) curve for CW Cas based on new medium-resolution spectroscopic observations. By simultaneously analyzing this RV curve along with multi-wavelength light curves, we derive the most reliable physical parameters of the binary system to date. Furthermore, we extend the \mbox{\it O\,--\,C} diagram by over half a century compared to previous studies to refine the orbital period investigation and improve constraints on the potential third body. Our findings, combined with Gaia astrometry, suggest that CW Cas may be a hierarchical quadruple system, possibly hosting a compact tertiary companion. Additionally, the evidence of possible spot activities and magnetic cycles within the system are shown as well.
%\textcolor{red}{this analysis is the first where we use spectral data to constrain mass ratio, unlike previous studies which were based only on photometry. Also this study uses archival photometry with time base $>$ 100 years so third body orbit is much better constrained.}}

\section{Photometric and Spectroscopic observations} \label{section2}

\subsection{\texorpdfstring{$BVR_{c}I_{c}$}{BVRcIc} band observation} \label{sec2.1}

New multiband photometric observations of the CW Cas were performed with the 60cm Carl Zeiss-600 Eastern Telescope (ET-60) on 2022 July 17-23 at Maidanak Astronomical Observatory (MAO) of the Ulugh Beg Astronomical Institute of the Uzbekistan Academy of Sciences \citep{Ehgamberdiev_2018NatAs...2..349E}. The optical system of the telescope is Cassegrain, and the focal length is 7500 mm. The FLI Series ProLine KAF-1001E CCD camera was used as the receiver. The reading noise and amplification coefficients of the CCD camera are 13.0  analog-digital units and 5.0e-, the effective field of view is $11\mathrm{'}.7 \times 11\mathrm{'}.7$ arcmin, and the pixel scale is $0\mathrm{''}.687$ pixel$^{-1}$. Image acquisition was done with MaxIm DL 6 version. The observations were made using the $BVR_{c}I_{c}$ filters of the Johnson-Cousins photometric system at exposure times of 90, 60, 30 and 30 seconds, respectively. In addition, $BVR_{c}I_{c}$ multiband observations of CW Cas were carried out on 2024 October 27 with the 1 m telescope at the Yunnan Observatories (YNOs) of the Chinese Academy of Sciences. One more observation was carried out on  2024 November 2 using the 70 cm Sino-Thai telescope. This telescope was located at the Lijiang Gaomeigu station of YNOs, which equipped an Andor DW936 2K CCD camera. 

\begin{figure*}[htb!]
	\begin{minipage}{0.5\linewidth}
		\center{\includegraphics[width=1\linewidth]{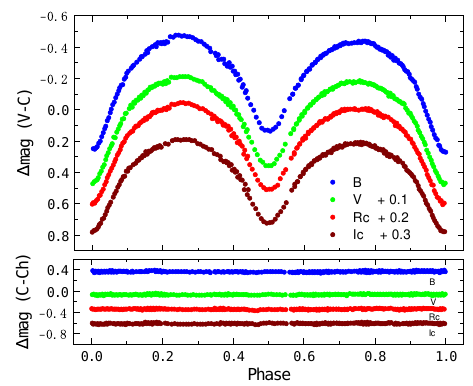}} \\
	\end{minipage}
	\hfill
	\begin{minipage}{0.5\linewidth}
		\center{\includegraphics[width=1\linewidth]{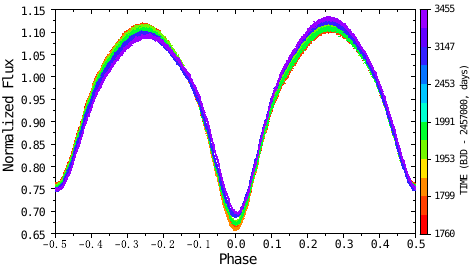}} \\
	\end{minipage}
	\caption{Left panel: The new $BVR_{c}I_{c}$ band light curves of CW Cas. %Top: The variable star (V) minus comparison star (C) for $BVR_{c}I_{c}$ bands. Bottom: comparison (C) minus the check star (CH). 
		Right panel: TESS light curves display variations in phase and flux diagram. The color bar column illustrates the changes across all five sectors.}
	\label{figure1}
\end{figure*}

CCD images were analyzed using aperture photometry with the DAOPHOT and APPHOT packages of Image Reduction and Analysis Facility (IRAF) software. This analysis included processes for flat-field, bias, and dark current correction. The variable star ("V"), comparison star ("C"), and check star ("Ch") are located in CW Cas (RA = $00^{h}45^{m}52.7^{s}$, Dec = $+63^{\circ} 05\mathrm{'} 08.4\mathrm{''}$) with $R_{mag}=11.26$, TYC 4020-1349 (RA = $00^{h}45^{m}09.7^{s}$, Dec = $+63^{\circ} 06\mathrm{'} 11.7\mathrm{''}$) with $R_{mag}=10.80$, TYC 4020-919 ([RA=$00^{h}45^{m}35.0^{s}$, Dec=$+63^{\circ}05\mathrm{'} 02.0\mathrm{''}$) with $R_{mag}=12.10$, respectively. The $\Delta$ magnitude of the variable star and the comparison star (V - C), as well as between the comparison star and the check star (C - Ch), are shown in the top and bottom panels, respectively, of the left panel of Figure \ref{figure1}. The standard deviations of the differential magnitude between the check star and the comparison star for each band are used to evaluate the errors of the data, which are 0.007 mag for $B$, 0.003 mag for V band and 0.0025 mag for $R_{c}, I_{c}$ band. The instrumental magnitudes were transformed into the standard photometric system using the extinction coefficients and transformation equations described by \citet{Artamonov_2010ARep...54.1019A} and \citet{Azimov_2023Atmos..14.1779A}.
%The coordinates of the variable star, the comparison star and the check star are listed in Table \ref{table1}. Because these two stars are very close to the target, their extinction conditions were closely similar. %\textcolor{red}{The average photometric errors for individual observations are 0.0033 mag in the $B$ band, 0.0028 mag in the $V$ band, and 0.0032 mag in the $R_{c}I_{c}$ band.}
%In addition,  $BVR_{c}I_{c}$ multiband observations of CW Cas were carried out in 27th October 2024 in the 1 m telescope at the Yunnan Observatories (YNOs) of the Chinese Academy of Sciences. And 2nd November 2024 in the 70 cm Sino-Thai telescope. This telescope located at the Lijiang Gaomeigu station of YNOs, which equipped an Andor DW936 2K CCD camera. Standard reductions of CCD images was performed using the DAOPHOT package of IRAF software, including bias and flat corrections and aperture photometry. From observations, 2 primary and 1 secondary light times obtained. 

\subsection{TESS Photomeric Data}\label{sec2.2}
The main purpose of the Transiting Exoplanet Survey Satellite (TESS) \citep{Ricker_2015JATIS...1a4003R} is to search for exoplanets.  In addition to executing its observation plan, TESS has provided valuable data on variable stars, including eclipsing binary stars. As a result of these observations, TESS has generated a substantial amount of high-precision photometric data for eclipsing binary systems. 
TESS provided observations for CW Cas in five sectors: sector-17 (BJD2458764-BJD2458789), sector-18 (BJD2458790-BJD2458815), sector-24 (BJD2458955-BJD2458982), sector-58 (BJD2459882-BJD2459910) and sector-78 (BJD2460434-BJD2460452), with an exposure time of 2 minutes. We utilized the Simple Aperture Photometry (SAP)\footnote{\url{https://heasarc.gsfc.nasa.gov/docs/tess/Aperture-Photometry-Tutorial.html}} 
light curves, which were downloaded from the Mikulski Archive for Space Telescopes (MAST)\footnote{\url{https://mast.stsci.edu/portal/Mashup/Clients/Mast/Portal.html}} database \citep{Stassun_2019AJ....158..138S}. All photometric data were converted to normalized flux using the Python $\texttt{Lightkurve}$ package \citep{Lightkurve_2018ascl.soft12013L}. 
%The light curves from five sectors, each represented in a different color, are displayed in the left panel of Figure \ref{figure1}. %The figure illustrates that the light curves exhibit significant variations at both the light maxima and minima within the 2019-2024 time domain.
We built a phase-flux diagram based on the time series data that were shifted the data at phase 0 which are shown in the right panel of Figure \ref{figure1}. In this figure, the color bar in the diagram represents the change over time.  The figure shows that the light curves show obvious variations at the light maxima and minima.

\subsection{Spectral observations}\label{section2.3}
To determine the atmospheric parameters of the CW Cas, we conducted spectral observations on 2024 October 22, using the Beijing Faint Object Spectrograph and Camera (BFOSC) mounted on $2.16$-meter telescope at the Xinglong station of the National Astronomical Observatories, Chinese Academy of Sciences (NAOC) \citep{Zhao_2018RAA....18..110Z}. We used E9+G10 grisms with a slit width of $1\mathrm{''}.6$ arcseconds, achieving a spectral resolution of approximately 13,000 over the range of 400–700 nm \citep{Fan_2016PASP_128k5005F}. The exposure time was set to 1800 seconds. We processed the observation images and extracted the spectra using IRAF. These spectra were taken at phases close to the secondary minimum ($\phi = $0.42, and 0.49) of the light curve.
%The normalized spectrum flux shown as black line right panel of Figure \ref{figure1}. 
To obtain stellar atmospheric parameters, we utilized the University of Lyon Spectroscopic Analysis Software (ULySS) package \citep{Koleva_2009AA_501.1269K}, which fits the full spectra against the model spectra generated by interpolation using the ELODIE library \citep{Prugniel_2001AA_369.1048P}. 
%The fitted spectra are shown as blue lines on the right panel of Figure \ref{figure1}. 
Based on two observations, the mean values of the stellar atmospheric parameters were derived as follows: effective temperature $T_{\rm{eff}}=5280(11)K$, surface gravity % If you want different parts in different styles:
log{\it g} = 3.69(2) cm s$^{-2}$, and metallicity [Fe/H]=0.60(1) dex. 

In order to obtain the radial velocity (RV) curve, We have carried out medium-resolution spectroscopic observations of the CW Cas on December 2024 29-30 and 2025 January 1-2, using the $2.4$ m telescope at the Thai National Observatory (TNO), National Astronomical Research Institute of Thailand (NARIT). 
%The telescope is equipped with an ARC 4K CCD camera and a medium-resolution spectrograph. 
The wavelength range of the spectrometer is 3800–9000 $\mathring{\mathrm A}$ with a resolution of $R\sim 18,000$. The raw spectra were processed with IRAF software, including spectral extraction, calibration, and normalization. Finally, nine normalized spectra and RVs were obtained. 
%The stellar atmospheric parameters and the RV are listed in Table \ref{table1}. 

\section{Orbital Period Analysis}\label{section3}

Orbital period variations are a widespread phenomenon in eclipsing binary systems, typically caused by loss of angular momentum, mass transfer, a third body, and magnetic activity.

To investigate variations in the orbital period, we collected light-time minima from various open-source databases and literature sources. From the public \mbox{\it O\,--\,C} Gateway\footnote{\url{http://var2.astro.cz/ocgate/}}, we obtained 389 minimum times, along with 5 new minima from additional literature. Using our multi-band photometric observations we calculated 10 new light minima (see \ref{sec2.1}). By quadratic parabolic fitting, we derived 596 new eclipse times (298 primary and 298 secondary) from the TESS data. Furthermore, using the same calculation method, we obtained 106 minima times in the {\it V-}band light curves between 2011 and 2024 from the American Association of Variable Star Observers (AAVSO\footnote{\url{https://www.aavso.org/databases}}) database. The All-Sky Automated Survey for Supernovae (ASAS-SN) \citep{Shappee_2014ApJ...788...48S, Christy_2023MNRAS.519.5271C} provided photometric observations in the $V$ and $g$ bands of CW Cas spanning 2012–2025. Since these data are not continuous, we employed template matching over phased time intervals of 150 days, using the phased TESS light curve from Sector 58 as a reference. This approach allowed us to extract an additional 26 minima from both the {\it V-} and {\it g-}band data. The Digital Access to a Sky Century at Harvard (DASCH\footnote{\url{https://dasch.cfa.harvard.edu/}}) project \citep{Grindlay_2012IAUS..285..243G} digitizes photographic plates covering the years 1889 to 1989. Although the DASCH data exhibit significant scatter and lower time resolution, with a mean photometric error of 0.18 mag, they are valuable for orbital period analysis due to their long time span and timing accuracy of approximately 0.0007 days. By applying quadratic parabolic fitting and reconstructing phases using data spanning more than one cycle, we derived a series of 116 eclipse times. Our further analysis shows that these eclipse timings obtained via the reconstruction method align well with eclipse timings from the literature derived using CCD/PE methods (\mbox{\it O\,--\,C} Gateway). This agreement confirms the reliability of the eclipse timings obtained through this method and demonstrates the significant contribution of the DASCH data to this study. All obtained eclipse times were converted to HJD format and are listed in the first column of Table \ref{table2}.

%\textcolor{red}{The Table \ref{table3} includes a eclipse times, the corresponding errors, the primary minimum (P) and secondary minimum (S), observation method (Visual/Photographic/Photoelectric/CCD), source of data and reference.}

\begin{table*}[htb!]
\footnotesize
	%\begin{threeparttable}
		\caption{The Light Minimum Times of CW Cas}
		\begin{tabular*}{\textwidth}{@{\extracolsep{\fill}}lcccccccc}
			\hline\hline
			\rm{Eclipse Timing} & \rm{Error} & \rm{Epoch} & $O-C$ & \rm{Primary(P)orSecondary(S)Min.} & \rm{Method} & \rm{Sources} &  \rm{Reference} \\
			\rm{(HJD)}     &    &    & \rm{(days)} &   Min   & {} & {}       &  {}        \\
			\hline
			2459782.41634 & 0.00012 & 22753   & -0.000750 & P & CCD & Maidanak 60cm & (1) \\
            2459783.37157 & 0.00017 & 22756   & -0.002100 & P & CCD & Maidanak 60cm & (1) \\
            2459784.32930 & 0.00016 & 22759   & -0.000960 & P & CCD & Maidanak 60cm & (1) \\
            2459778.43088 & 0.00018 & 22740.5 & -0.000430 & S & CCD & Maidanak 60cm & (1) \\
            2459779.38869 & 0.00007 & 22743.5 &  0.000790 & S & CCD & Maidanak 60cm & (1) \\
            2459780.34353 & 0.00016 & 22746.5 & -0.000950 & S & CCD & Maidanak 60cm & (1) \\
            2459782.25690 & 0.00009 & 22752.5 & -0.000760 & S & CCD & Maidanak 60cm & (1) \\
            2460611.12790 & 0.00010 & 25352   & -0.011520 & P & CCD & YNOs 1 m       & (1) \\
            2460618.14262 & 0.00022 & 25374   & -0.011770 & P & CCD & YNOs 70 cm   & (1) \\
            2460617.02710 & 0.00013 & 25370.5 & -0.011270 & S & CCD & YNOs 70 cm   & (1) \\
            2451850.1682  & 0.0001  & -2124   &  0.08109  & P & CCD & \citet{Jiang_2010PASJ...62..457J} & (2) \\
			\hline
		\end{tabular*}
		\label{table2}
		\begin{tablenotes}
			\item References: (1) This paper; (2) \cite{Jiang_2010PASJ...62..457J}; (3) \mbox{\it O\,--\,C} Gateway. 
            %(3) DASCH;  (4) TESS; (5) ASAS-SN; (6) AAVSO database;
            \item (This table is available in its entirety in machine-readable format) 
		\end{tablenotes}
	%\end{threeparttable}
\end{table*}

\begin{figure*}[htb!]
	\centering
	\includegraphics[scale=1.0]{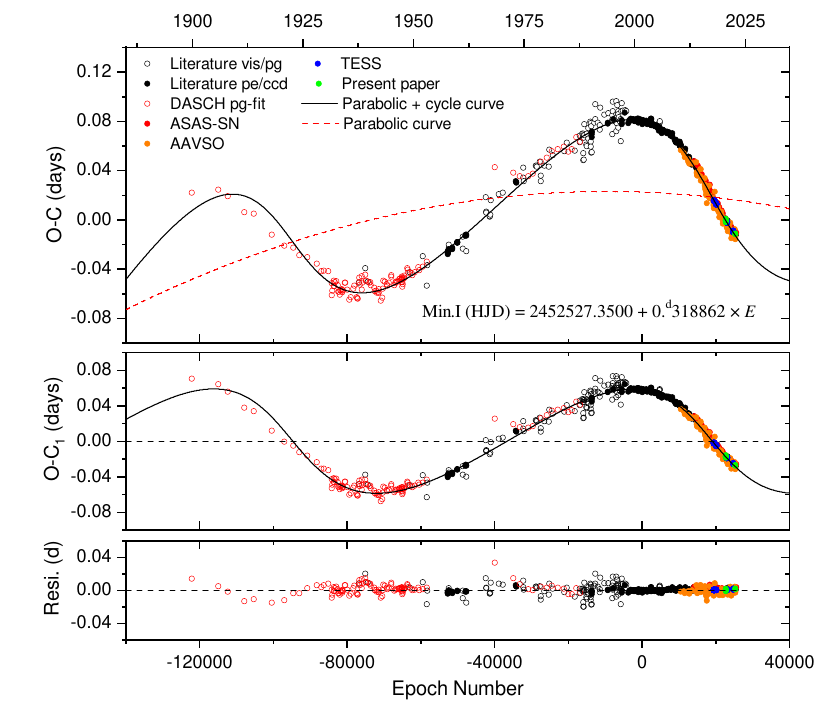}
	\caption{The \mbox{\it O\,--\,C} diagram of CW Cas fitted with an eccentric orbit. Upper panel: The red dashed line are parabolic changes. The black solid line refer to a downward parabolic plus cyclic variation. Middle panel: \mbox{\it O\,--\,C}$_{1}$ diagram after subtracting the parabolic variation. Bottom panel: Final residuals after subtracting parabolic and periodic variations.
		\label{figure2}}
\end{figure*}

Using the same linear ephemeris as the \mbox{\it O\,--\,C} Gateway,
%The O-C value was calculated according to the following linear ephemeris formula:
\begin{equation}
	\rm{Min.I} (HJD) = 2452527.3500 + 0^{d}.318862 \times \it{E}.
	\label{equation1}
\end{equation}
the \mbox{\it O\,--\,C} values were calculated and are listed in Column 4 of Table \ref{table2}. The corresponding \mbox{\it O\,--\,C} curve is displayed in the upper panel of Figure \ref{figure2} against the epoch number $E$. Our newly determined times of minimum light extend the \mbox{\it O\,--\,C} diagram by over half a century comparing to previous studies. Based on this \mbox{\it O\,--\,C} curve spanning 125 years, we could improve the orbital period investigation. 
% where the red and black circles are the PG  minima from DASCH and literature, the black dots are the PE/CCD minima from the literature, the green dots are TESS minima, and the blue dots are the our observation in 2022 at MAO. 

%To describe the \mbox{\it O\,--\,C} curve well, a cyclic variation and longterm periodical oscillation variation superposed used on the downward parabolic with a periodic oscillation by Equation \ref{eq:a2}. 
%According to the \mbox{\it O\,--\,C} diagram, it is clearly seen that there is a parabolic variation, and possibly superimposed by a cyclic oscillation.

In \mbox{\it O\,--\,C} curves, it is clear that a periodic oscillation exists, which may be caused by LTTE due to a third body as suggested by the previous authors. \citep{LiQian_2010AJ....139.2639L}. 
%Therefore, we attempted to fit the \mbox{\it O\,--\,C} values with a downward parabolic curve with a periodic oscillation. 
Here, we considering the eccentric case by using the model of light-time effect to analysis such variation \citep{Irwin_1952ApJ...116..211I}, \mbox{\it O\,--\,C} diagram was described by the following equation:
%The eccentric fitting formula used in this work is shown in Equation \ref{equation2}, proposed by \citep{Irwin_1952ApJ...116..211I}. 
\begin{equation}
	\begin{split} 
		O - C & = \Delta T_{0} + \Delta P_{0} \times E + \frac{\beta}{2} \times E^{2} \\ 
		& + A \Biggl[(1 - e^{2}_{3})\:\frac{\sin(\nu_{3} + \omega_{12})}{1 + e_{3}\:\cos(\nu_{3})}  +  e_{3}\:\sin \omega_{12} \Biggr] \\
		& =  \Delta T_{0} + \Delta P_{0} \times E + \frac{\beta}{2} \times E^{2} \\ 
		&+ A \Biggl[ \sqrt{1 - e^{2}_{3}} \sin E^{*} \cos \omega_{12} + \cos E^{*} \sin \omega_{12}\Biggr]
	\end{split} 
	\label{equation2}
\end{equation}
where $E$ is epoch number, $\Delta T_{0}$, $\Delta P_{0}$ are the correction value of initial epoch and orbital period of binary system, $\beta$ is the long-term change of the orbital period (d cycle$^{-1}$), $A=a_{12}\sin i_{3}/c$ and $a_{12}\sin i_{3}$ is the projected semi-major axis given in days, $e_{3}$ is the eccentricity of supposed third body, $\nu_{3}$ is the true anomaly, $\omega_{12}$ is the longitude of periastron passage in the plane of orbit, and $E^{*}$ is the eccentric anomaly \citep{LiQian_2010AJ....139.2639L}.

The Kepler equation provides the connection between the eccentric anomaly $(E^{*})$ and the observed times of the light minimum,
\begin{equation}
	M^{*} = E^{*} - e_{3} \sin E^{*} = \frac{2\pi}{P_{3}}(t-T_{3})
	\label{equation3}
\end{equation} 
where $M^{*}$ is the mean anomaly, $T_{3}$ is the time of periastron passage, $P_{3}$ is the period of a supposed third body, $e_{3}$ is the eccentricity of the third body orbit and $t$ is the observed times of light minimum. It is clear from equations (\ref{equation2}) and (\ref{equation3}) that we could determine five parameters ($P_{3}$, $T_{3}$, $A$, $\omega_{12}$, $e_{3}$) that fit the trend of the \mbox{\it O\,--\,C}. For this case, the eccentric anomaly $E^{*}$ is approximately in a Bessel series \citep{LiQian_2010AJ....139.2639L}. Then the Levenberg-Marquardt method is adopted using equations (\ref{equation2}) and (\ref{equation3}) for a nonlinear fit of the \mbox{\it O\,--\,C} diagram. 

We find that the orbital period of CW Cas exhibits a cyclic variation with a period of $P_{3}=99.4(6)$ yr and a semiamplitude of $A=0.0628(5)$ d, superimposed on a long-term period decrease at a rate of $\beta=-1.12(8) \times 10^{-11}$ d cycle$^{-1}$. The observed cyclic variation could be attributed to the light-travel-time effect (LTTE) caused by the motion of a third body\citep{LiaoQian_2021MNRAS.508.6111L}, and the eccentricity of the tertiary orbit is $e_{3}=0.352(8)$. The results of the \mbox{\it O\,--\,C} analysis fitted using Equations (\ref{equation2}) and (\ref{equation3}) are summarized in the upper part of Table \ref{table3}, and the corresponding \mbox{\it O\,--\,C} diagrams are shown in Figure \ref{figure2}.The top panel of Figure \ref{figure2} displays a downward parabolic trend with superimposed periodic oscillations, where the red dashed line indicates the parabolic component and the black solid line represents the combined parabolic and cyclic variations. The middle and bottom panels show the \mbox{\it O\,--\,C} diagram after removal of the parabolic trend and the final residuals, respectively. %Open circles denote visual (Vis) and photographic (PG) data, while filled symbols represent photoelectric (PE) and CCD data.}}

%The results corresponding to the fitting of the \mbox{\it O\,--\,C} analysis using equations (\ref{equation2}) and (\ref{equation3}) are presented in the upper part of Table \ref{table3}. The calculated \mbox{\it O\,--\,C} diagrams are shown in Figure \ref{figure2}. As shown in the top panel of Figure \ref{figure2}, the \mbox{\it O\,--\,C} diagram exhibits a downward parabola with a periodic oscillation, where the red dashed line represents the trend of the parabolic curve, while the black solid line represents the combined parabolic variation and plus the cyclic variations. The middle panel of Figure \ref{figure2}, represents the \mbox{\it O\,--\,C} diagram after subtracting the parabolic trend, and the final residuals after all trends have been removed are displayed in the bottom panel of Figure \ref{figure2}.  The open circles represent visual (Vis) and photographic (PG) data, while all dots represent photoelectric (PE) and CCD data.

\begin{table}[htb!]
\footnotesize
	\centering
	\caption{Orbital parameters of the third body in CW Cas.}
	%\begin{tabular*}{\columnwidth}{@{\extracolsep{\fill}}lcc}
    %\begin{tabular}{p{0.52\columnwidth}cc}
    \resizebox{\columnwidth}{!}{%
    \begin{tabular}{lcc}
		\hline\hline
		Parameters & Value & Unit \\
		\hline
		Revised epoch, $\Delta T_{0}+T_{0}$ (HJD0) & $2452527.37234(57)$ & days \\
		Revised period, $\Delta P_{0}+P_{0}$ & $0.31886189(3)$ & day \\
		Long-term change of the orbital period, $\beta$ & $-1.12(8)\times10^{-11}$ & day cycle$^{-1}$ \\
        %period, $\beta$ &  &  \\
		Semi-amplitude, $A$ & $0.0628(5)$ & day \\
		Projected semi-major axis, $a_{12}\sin i_{3}$ & $10.87(8)$ & a.u. \\
		Eccentricity, $e_{3}$ & $0.352(8)$ & \\
		Longitude of periastron, $\omega$ & $174.5(1.7)$ & degree \\
		Orbital period, $P_{3}$ & $99.4(6)$ & year  \\
		Mass function, $f(m)$ & $0.130(3)$ & $M_{\odot}$ \\
		Time of periastron passage, $T_{3}$ & $2458039(90)$ & days \\
		Semi-major axis, $a_{\rm{3max}}$ & $12.0(2)$ & a.u. \\
		Projected mass, $m_{\rm{3min}}$ & $0.91(1)$ & $M_{\odot}$ \\
		\hline
	\end{tabular}
    }
	\label{table3}
\end{table}

%Where the red circles represents the minima from DASCH, the black circle and dot are indicate the Vis and PE/CCD minima from the literature. The blue dots represent the minima from our $BVR_{c}I_{c}$ photometric observation, while the green dots represent TESS minima. The middle panel of Figure \ref{figure5} represents the \mbox{\it O\,--\,C} diagram after subtracting the parabolic trend, and the bottom panel shows the final residuals after removing all trends.  %The corresponding results of Equations \ref{eq2} and \ref{eq3} are listed in the Table \ref{tbl4}. 

%The eclipse timings obtained using the reconstructing method; align with the eclipse timing from literature obtained using CCD/PE (O-C Gateway). This alignment proves the reliability of the eclipse timing obtained with this method and demonstrates the significant contribution of the DASCH data to this study.

\section{Photometric and spectroscopic study}\label{section4}

\subsection{Light curve variation}\label{section4.1}
 
%To investigate the relative variations in the light curve, we build a phase-flux diagram based on the time-series data that were shifted in phase 0 and are shown in the right panel Figure \ref{figure1}.

\begin{figure*}[htb!]
\centering
\includegraphics[scale=1.0]{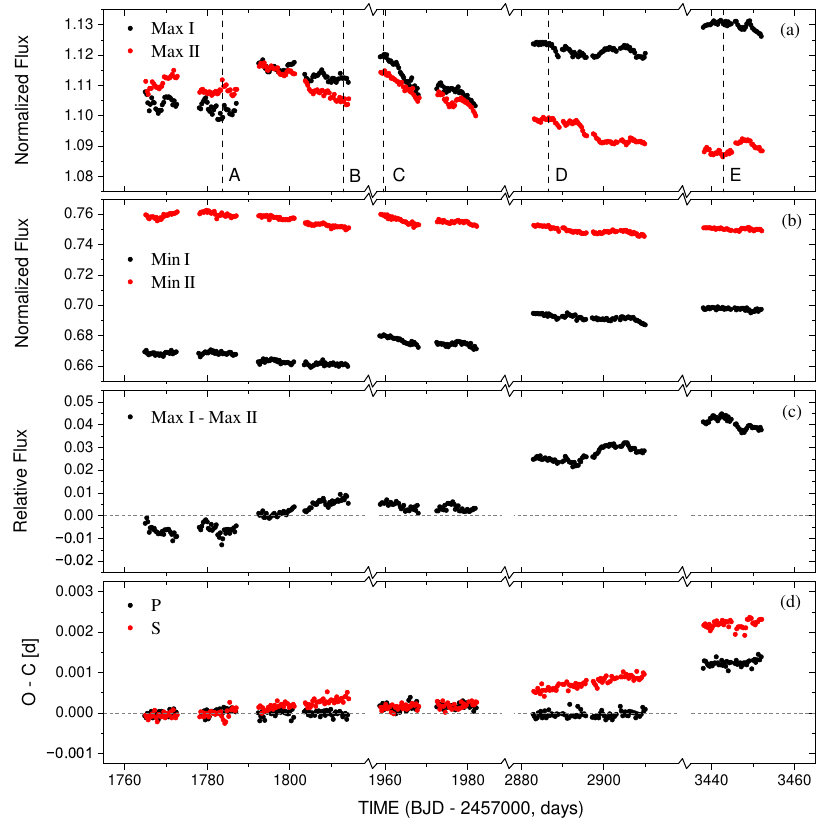}
\caption{Panel (a): Max\ I and Max\ II changes relationship diagram, $A\sim E$ represent five different stages. Panel (b): Min I and Min II change relationship diagram. Panel (c): Max I - Max II change relationship diagram. Panel (d): \mbox{\it O\,--\,C} diagram for TESS light curves. 
\label{figure4}}
\end{figure*}

By comparing the light curves from different TESS sectors, we found asymmetries in both maxima. To investigate the relative variations in the light curve, we constructed a phase–flux diagram from the time-series data, shifted to phase zero, which is shown in the right panel of Figure \ref{figure1}.
The color bar in the diagram represents the change over time. In this figure, one can see that the light curves of the CW Cas change smoothly. It was found that the O'Connell effect switched from positive to negative. The positive O'Connell effect states that the luminosity at phase $-0.25$ is greater than that at phase $0.25$, and the initial flux of the Max-II at phase $-0.25$ is bigger than the flux of the Max-I at phase $0.25$, but later it was reversed. As shown in the right panel of Figure \ref{figure1}, the eclipse depth changes in phase 0.0 (Min-I). The initial Min-I depth in BJD-1760 is deeper than in BJD-3435. The variations in the different light levels with respect to BJD time in days (BJD - 2457000) are shown in Figure \ref{figure4}. Panel (a) displays the variations of Max-I and Max-II; panel (b) shows the variations of Min-I and Min-II; panel (c) illustrates the difference between Max-I and Max-II; and panel (d) presents the \mbox{\it O\,--\,C} curve based on the newly calculated orbital period for TESS. In panel (a), five distinct time intervals (stages $A \sim E$) represent different TESS sectors are marked by dashed vertical lines, which are used for WD-code modeling with a dark spot. Panels (a) and (b) reveal that the primary and secondary maxima and minima, as well as the \mbox{\it O\,--\,C} curve in panel (d), are anticorrelated over the same time interval. This behavior may be due to spot migration on the stellar surface, as suggested by \citet{KalimerisO-C_spot2002AA...387..969K,Tran_2013ApJ...774...81T,Balaji_2015MNRAS.448..429B}. Therefore, it implied that CW Cas should be a binary with magnetic activity, and may have dark spots on the surface of its components). %It is worth noting that all four panels show a significant negative correlation.

\subsection{Radial velocity curve and Light-curve analysis}\label{section4.2}

\begin{figure*}[htb!]
	\begin{minipage}{0.5\linewidth}
		\center{\includegraphics[width=1\linewidth]{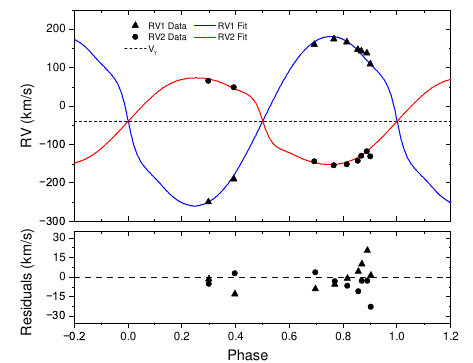}} \\
	\end{minipage}
	\hfill
	\begin{minipage}{0.5\linewidth}
		\center{\includegraphics[width=1\linewidth]{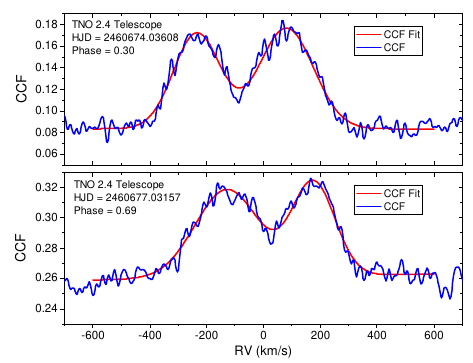}} \\
	\end{minipage}
	\caption{Left panel: The RV curves of CW Cas. Triangles and circles represent components with velocities V1 and V2, respectively. Our model included Rossiter-McLaughlin effect \citep{Rossiter_1924ApJ....60...15R, McLaughlin_1924ApJ....60...22M}. 
		Right panel: CCF profile of the spectrum. The solid blue line indicates the CCF profile. The red solid line represents the double Gaussian fit.}
	\label{figure3}
\end{figure*}

Given that CW Cas is a magnetically active, partially eclipsing binary, the determination of accurate stellar parameters necessitates the combination analysis of both radial velocity (RV) variations and photometric light-curve data. Based on the medium-resolution spectra obtained by the TNO 2.4m telescope, we can derived the radial velocities of the two components of CW Cas by the cross-correlation function (CCF) method. One of the CCF profile is shown in the right panel of Figure \ref{figure3}. Two components of this system can be seen clearly in this figure. The stellar atmospheric parameters and RVs are listed in Table \ref{table4}. The phase-folded RV curves are presented in the left panel of Figure \ref{figure3}. Based on this RV curve, the velocity amplitudes $K_{1.2}=227.4(5.1), 120.6(5.7)$ km s$^{-1}$ and the center-of-mass velocity $V_{\gamma}=-38.95(3.48)$ km s$^{-1}$ are derived. Thus, we derived the spectroscopic mass ratio $q=M_{2}/M_{1}=1.88(9)$.

%At the same time, based on the orbital period analysis, we hypothesize the existence of a third body with a minimum mass of $0.91 M_{\odot}$, located at a maximum distance of 12 AU from the primary components of the binary system (see Table \ref{table3} in Sec. \ref{section2}). Thus, it can be assumed that the spectrum of this third body can be observed simultaneously with the spectrum of the binary system, as demonstrated for {\bf J04+25} \citep{Kovalev_2025MNRAS.539.3830K} and NY Boo \citep{Meng_2023ApJ...954..111M}.

\begin{table*}[htb!]
%\footnotesize
	\centering
	\caption{The Information of the Xinglong 2.16 m and TNO 2.4 m spectra}
    %{\small
	\begin{tabular*}{\textwidth}{@{\extracolsep{\fill}}ccccccccc}
		\hline\hline
		\multicolumn{9}{c}{Xinglong station 2.16 m telescope} \\
		\hline
		Date of Observation (UTC) & HJD & Phase & $T_{\rm{eff}}$ & log (\sl{g}) & Fe/H & Type & RV$_{1}$   & RV$_{2}$ \\
		(YYYY.MM.DD HH:MM:SS)     &     &       & (K)       &            &      &      & (km s$^{-1}$) & (km s$^{-1}$) \\
		\hline
		2024.10.22 15:38:38 & 2460606.15183 & 0.424 & 5318(11) & 3.79(2) & -0.502(1)  & G8 &  & \\
		2024.10.22 16:09:46 & 2460606.17345 & 0.492 & 5242(12) & 3.579(3) & -0.706(1)  & G8 &  & \\
		\hline
		\multicolumn{9}{c}{Thailand National Observatory 2.4 m Telescope} \\
		\hline
		%DATE-OBS  & HJD & Phase & $T_{\rm{eff}}$ & log (\sl{g}) & Fe/H & Type & RV$_{1}$   & RV$_{2}$ \\
		%(UTC)     &     &       & (K)  &           &      &      & (km s$^{-1}$) & (km s$^{-1}$) \\
		%\hline 
		2024.12.29 12:48:55 & 2460674.03608 & 0.298 & - & - & -  & - & -249.49(1.38) &   65.20(1.39)  \\
        2024.12.29 13:32:34 & 2460674.06639 & 0.393 & - & - & -  & - & -188.87(1.33) &   48.56(1.45)  \\
        2024.12.30 15:42:40 & 2460675.15668 & 0.813 & - & - & -  & - & 166.40(1.03)  & -151.47(1.26)  \\
        2024.12.30 16:01:32 & 2460675.16978 & 0.854 & - & - & -  & - & 147.08(1.56)  & -142.88(1.84)  \\
        2024.12.30 16:16:56 & 2460675.18048 & 0.887 & - & - & -  & - & 137.56(1.85)  & -118.42(2.48)  \\
        2025.01.01 13:15:48 & 2460677.05459 & 0.765 & - & - & -  & - & 174.99(1.18)  & -154.63(1.45)  \\
        2025.01.01 12:42:39 & 2460677.03157 & 0.693 & - & - & -  & - & 160.08(1.44)  & -143.91(1.83)  \\
        2025.01.02 12:59:50 & 2460678.04345 & 0.866 & - & - & -  & - & 143.45(1.24)  & -129.52(1.54)  \\
        2025.01.02 13:15:14 & 2460678.05415 & 0.900 & - & - & -  & - & 109.02(3.32)  & -130.92(3.26)  \\
		\hline
	\end{tabular*}%}
	\label{table4}
\end{table*}

\begin{figure*}
	\begin{minipage}{0.5\linewidth}
		\center{\includegraphics[width=1\linewidth]{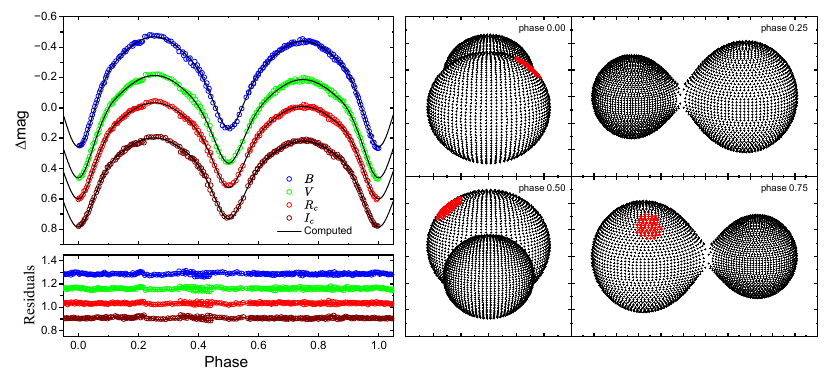}} \\
	\end{minipage}
	\hfill
	\begin{minipage}{0.5\linewidth}
		\center{\includegraphics[width=1\linewidth]{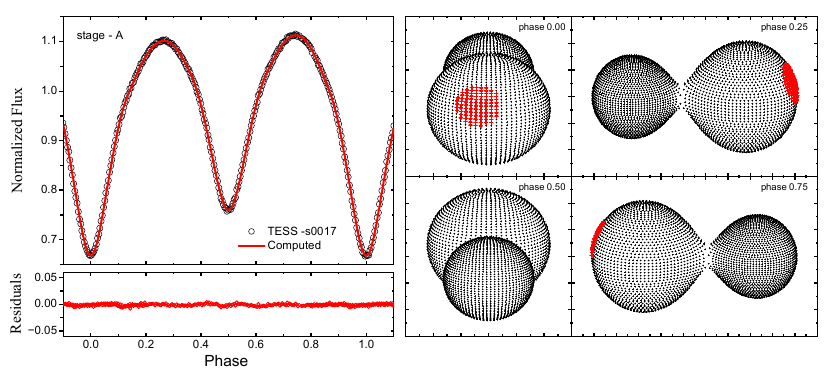}} \\
	\end{minipage}
	\hfill
	\begin{minipage}{0.5\linewidth}
		\center{\includegraphics[width=1\linewidth]{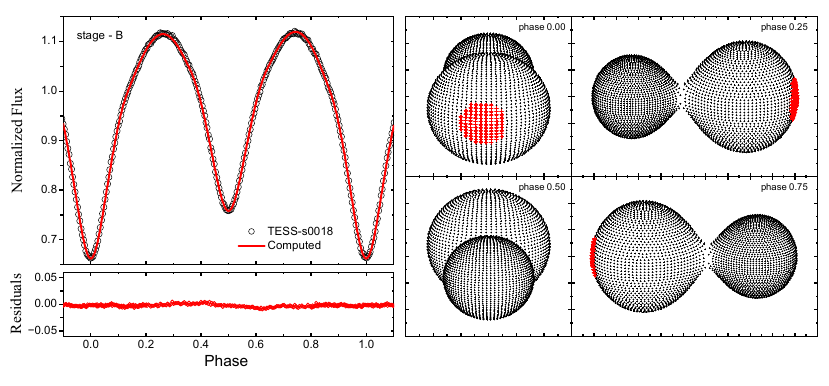}} \\
	\end{minipage}
	\hfill
	\begin{minipage}{0.5\linewidth}
		\center{\includegraphics[width=1\linewidth]{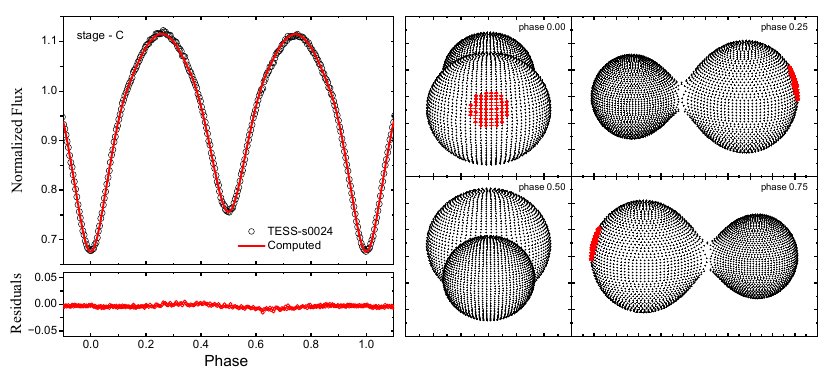}} \\
	\end{minipage}
	\hfill
	\begin{minipage}{0.5\linewidth}
		\center{\includegraphics[width=1\linewidth]{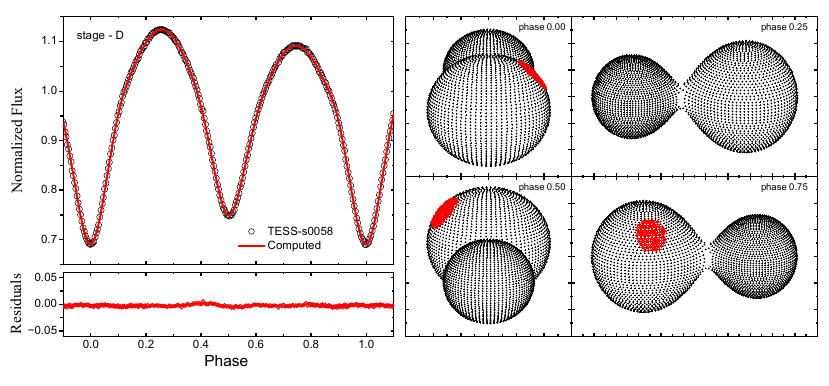}} \\
	\end{minipage}
	\hfill
	\begin{minipage}{0.5\linewidth}
		\center{\includegraphics[width=1\linewidth]{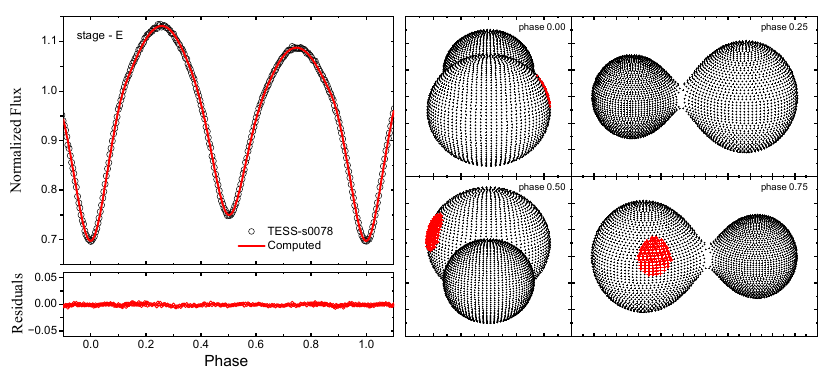}} \\
	\end{minipage}
	\caption{\rm{W-D theoretical light curves are compared with the observed light curves. W-D theoretical light curves and geometric configurations are presented for the BVRI light curves and five different stages ($A\sim E$) of the TESS light curves. The spot area is marked in red color.}}
	\label{figure5}
\end{figure*}

In order to obtain the reliable parameters of CW Cas, we simultaneously analyze this radial velocity (RV) curve and light curves by employing the Wilson–Devinney (W-D) program. \citep{Wilson_1971ApJ...166..605W, Wilson_2012AJ....144...73W, WilsonHamme_2014ApJ...780..151W}. Before running the W-D program, we fixed some parameters: The effective temperature of star 1 was fixed at 5280 K according to the stellar atmospheric parameters listed in Table \ref{table4}. The gravity darkening coefficient $g_{1}=0.32$,  $g_{2}=0.32$ and bolometric albedo $A_{1}=0.5$, $A_{2}=0.5$ were adopted for the convective envelope \citep{Lucy_1967ZA.....65...89L, Rucinski_1969AcA....19..245R}. The bolometric and bandpass limb-darkening coefficients were chosen from the table of \citep{vanHamme_1993AJ....106.2096V}. The mass ratio was fixed at 1.88. The adjustable parameters were: the orbital inclination $i$; the effective temperature of the star 2 $T_{2}$; the monochromatic light of star 1 $L_{1}$, the dimensionless potential of star 1 ($\Omega_{1} = \Omega_{2}$) (We selected an overcontact binary configuration after set of trial simulations). To account for the asymmetries present in both our $BVR_{c}I_{c}$ light curves and TESS light curves, we implemented a spot model with spot center longitude $(\theta)$, spot center latitude $(\phi)$ and spot angular radius$(r_{s})$ treated as adjustable parameters as well. Final converged solutions were successfully derived for the $BVR_{c}I_{c}$ light curves and each of the TESS light curves, corresponding to the five distinct stages (stage $A\sim E$). The results of the photometric solutions are listed in Table \ref{table5}. Figure \ref{figure5} shows all the theoretical light curves together with their geometric configuration at the 0.00, 0.25, 0.50 and 0.75 phases, with red areas representing the spot. The theoretical model shows good agreement with the observations, with the dark spot’s position and size differing between the $BVR_{c}I_{c}$ and TESS light curves, which cover a sufficiently large time span. Our solution suggests that the asymmetry of the light curve can be satisfactorily explained by the presence of a dark spot on the more massive star. As the analysis of the \mbox{\it O\,--\,C} curve suggested the existence of the third body, we tested for a third-light contribution by including $l_{3}$ as a free parameter in all light-curve fits, but found no significant improvement, indicating that its effect is negligible.

\begin{table*}[htb!]
    \centering
	\caption{Photometric solutions for CW Cas.} 
	\begin{tabular*}{\textwidth}{@{\extracolsep{\fill}}lcccccc}
		\hline\hline
		Parameters	&	$BVR_{c}I_{c}$	& TESS-s0017 &	TESS-s0018	&	TESS-s0024	&	TESS-s0058	&	TESS-s0078	\\
			 		&			        & A          &	B	        &	C	        &	D	        &	E \\
		\hline
		$q$ & 1.88 & 1.88 & 1.88 & 1.88 & 1.88 & 1.88 \\
		$i$ (degree) & 76.79(27)	& 75.35(98) & 75.92(4) & 75.63(5) & 73.50(12) & 73.69(5) \\
		$T_{1}$ (K)	& 5280 & 5280 & 5280 & 5280 & 5280 & 5280 \\
		$T_{2}$ (K)	& 5003(14) & 5064(13) & 5068(9) & 5060(8) & 5033(1) & 5022(2) \\
		$\omega_{1}=\omega_{2}$	& 5.0055(7) & 5.0041(41) & 5.0041(43) & 4.9664(67) & 4.9956(41) & 4.9968(17) \\
		$L_{1}/(L_{1}+L_{2}) (B)$     & 0.4393(17) & ... & ... & ... & ... & ... \\
		$L_{2}/(L_{1}+L_{2}) (B)$     & 0.5607(17) & ... & ... & ... & ... & ... \\
		$L_{1}/(L_{1}+L_{2}) (V)$     & 0.4250(17) & ... & ... & ... & ... & ... \\
		$L_{2}/(L_{1}+L_{2}) (V)$     & 0.5750(17) & ... & ... & ... & ... & ... \\
		$L_{1}/(L_{1}+L_{2}) (R_{c})$ & 0.4106(18) & ... & ... & ... & ... & ... \\
		$L_{2}/(L_{1}+L_{2}) (R_{c})$ & 0.5894(18) & ... & ... & ... & ... & ... \\
		$L_{1}/(L_{1}+L_{2}) (I_{c})$ & 0.3943(19) & ... & ... & ... & ... & ... \\
		$L_{2}/(L_{1}+L_{2}) (I_{c})$ & 0.6057(19) & ... & ... & ... & ... & ... \\
		$L_{1}/(L_{1}+L_{2}) (TESS)$  & ... & 0.4008(1) & 0.3999(1) & 0.40288(92) & 0.43480(8) & 0.43027(54) \\
		$L_{2}/(L_{1}+L_{2}) (TESS)$  & ... & 0.5992(1) & 0.6001(1) & 0.59712(92) & 0.56520(8) & 0.56973(54) \\
		$r_{1}$(pole)   & 0.31198(64) & 0.31199(38) & 0.31208(40) & 0.31533(63) & 0.31285(38) & 0.31280(15) \\
		$r_{1}$(side)	& 0.32691(78) & 0.32692(46) & 0.32703(48) & 0.33097(77) & 0.32796(46) & 0.32790(19) \\
		$r_{1}$(back)	& 0.36430(12) & 0.36432(73) & 0.36449(77) & 0.37090(13) & 0.36598(74) & 0.36588(30) \\
        $r_{2}$(pole)	& 0.41661(62) & 0.41673(37) & 0.41673(39) & 0.42013(61) & 0.41749(37) & 0.41739(15) \\
        $r_{2}$(side)	& 0.44354(81) & 0.44370(47) & 0.44370(50) & 0.44813(79) & 0.44469(48) & 0.44456(19) \\
        $r_{2}$(back)	& 0.47490(11) & 0.47507(64) & 0.47507(67) & 0.48100(11) & 0.47640(64) & 0.47621(26) \\
        $f (\%)$        & 14.00(12)   & 14.18(69)   & 14.18(73)   & 20.6(11)    & 15.62(69)   & 15.42(28)   \\
		Co-latitude, $\varphi$(radian) & 0.76298 & 1.26428 & 1.61265 & 1.34059 & 0.90339 & 1.29536 \\
		Longitude, $\theta$(radian)    & 4.83186 & 2.94823 & 3.05133 & 3.14792 & 4.87950 & 4.92833 \\
		Radius, $r_{s}$(radian)        & 0.24648 & 0.31785 & 0.32635 & 0.25524 & 0.28524 & 0.30686 \\
		Temp.factor $T_{s}/T$	       & 0.76095 & 0.76095 & 0.76095 & 0.76095 & 0.76095 & 0.76095 \\
		\hline 
	\end{tabular*}
	\label{table5}
\end{table*}

\section{DISCUSSIONS AND CONCLUSIONS}\label{section5}

\subsection{Absolute parameters}\label{section5.1}

%\textbf{The RV of CW Cas was determined using the CCF method applied to spectroscopic data obtained with the TNO 2.4 m telescope. The spectroscopic mass ratio was found to be $q = 1.88(9)$, and the systemic velocity was measured as $V_{\gamma} = -38.95(3.48)$ km s$^{-1}$ from the RV curves. By combining multi-color light curves with the radial velocity curve, we obtained a reliable solution using the WD program. This solution is also consistent across different stages (stages $A\sim E$) of the TESS light curves, with an average fill-out factor of 15\%. The distortion and asymmetry of the light curves were well fitted by adding a dark spot to the primary component. These phenomena indicate that CW Cas is an active binary system.} 

Based on photometric and spectral elements, the masses of the two components are calculated as $M_{1}=0.52(4)M_{\odot}$ and $M_{2}=0.98(6)M_{\odot}$ respectively \citep{Kopal_1959cbs..book.....K}. Combining Kepler's third law and the Stefan-Boltzmann law, the separation of the two components, radius, and luminosity can be derived \citep{LiKai_2021AJ....162...13L, Xu_2022RAA....22c5024X}. All absolute parameters of CW Cas are listed in Table \ref{table6}. Based on the results of our analysis, we conclude that CW Cas is an active W-subtype shallow-contact binary ($f = 15\%$) with a mass ratio of $q = 1.88$. CW Cas has been extensively studied by previous researchers. To facilitate comparison, we have compiled all existing results on this system in Table \ref{table7}. By employing a combined analysis of the radial velocity curve and light curves, our study derives the most accurate and reliable set of parameters for CW Cas to date.% together with the results of previous studies.

\begin{table}[htb!]
	\centering
	\caption{Absolute parameters of CW Cas}\label{table6}
    %\resizebox{\columnwidth}{!}{%
    %\begin{tabularx}{\columnwidth}{l>{\centering\arraybackslash}X>{\centering\arraybackslash}X}
	\begin{tabular*}{\columnwidth}{@{\extracolsep{\fill}}lcc}
		\hline \hline 
		Parameters & Star 1 & Star 2 \\
		\hline
		Mass $M_{\odot}$ & 0.52(4) & 0.98(6) \\
		Radius $R_{\odot}$ & 0.75(1) & 1.00(2) \\
		Luminosity $L_{\odot}$ & 0.39(1) & 0.56(2) \\
		Semi-major axis $R_{\odot}$ & \multicolumn{2}{c}{2.25(5)} \\
		\hline
	\end{tabular*}
    %\end{tabularx}
\end{table}

%\begin{equation}
%	K_{1,2} = \frac{2\pi a_{1,2} \rm{sin} \it{i}}{P \sqrt{1 - e^{2}}}
%	\label{equation4}
%\end{equation}
%\begin{equation}
%	M_{1,2} (M_{\odot}) = 1.0385 \times 10^{-7} \frac{(K_{1} + K_{2})^2 K_{1,2} P}{\rm{sin^{3}} \it{i}}
%	\label{equation4}
%\end{equation}

\begin{figure}[htb!]
	\centering
	\includegraphics[scale=0.55]{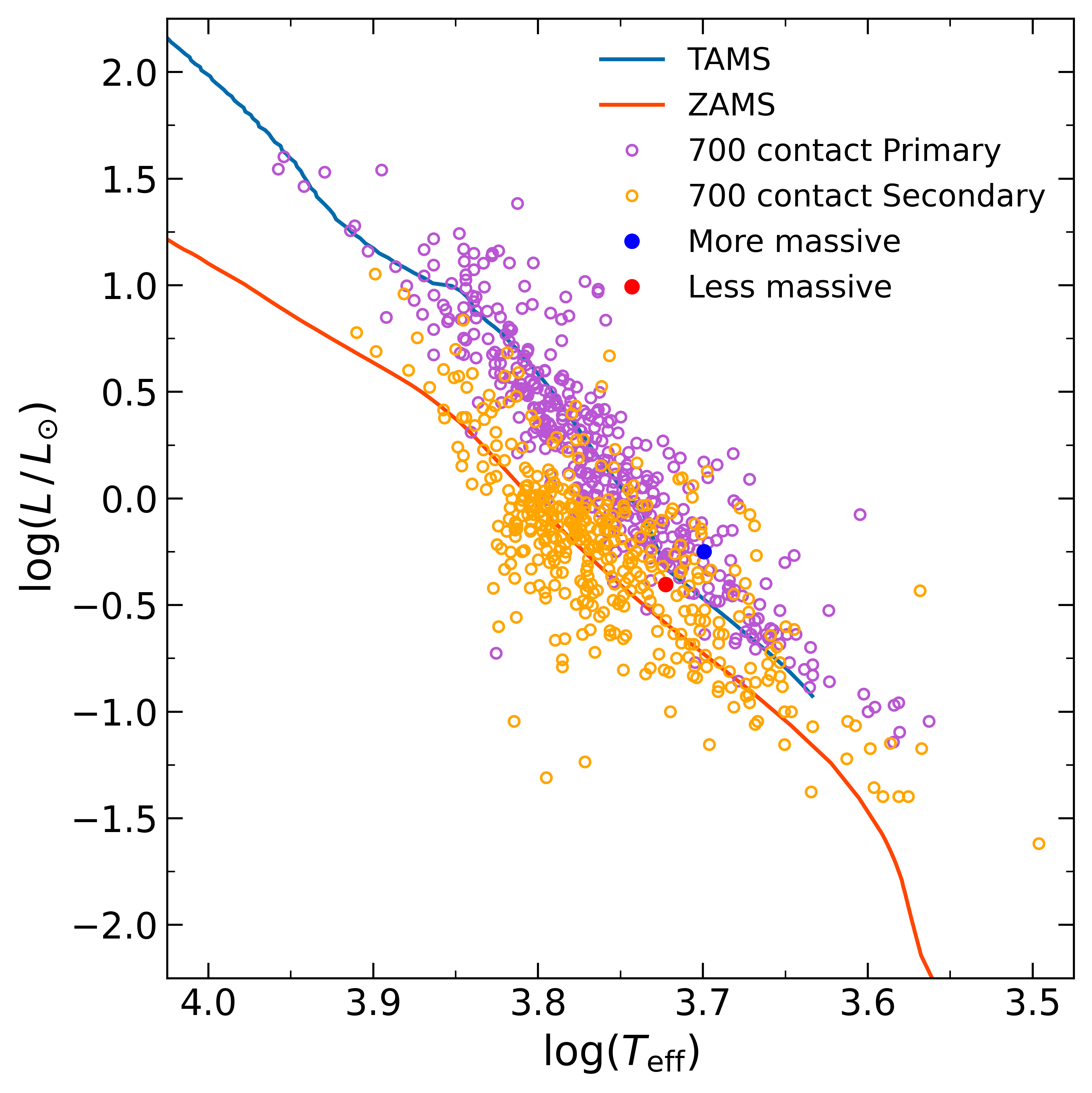}
	\caption{The position of the primary and secondary components of the CW Cas in the Hertzsprung-Russell diagram.
		\label{figure6}}
\end{figure}

In order to know the evolutionary stage of CW Cas, we have plotted it in the Hertzsprung-Russell diagram together with other contact binaries collected by \citep{Latkovic_2021ApJS..254...10L} in Figure \ref{figure6}. In this Figure, primary stars are more massive ones and the secondaries are less massive ones. The more massive component of CW Cas is marked in a blue dot, which is located upper in the terminal-age main sequence (TAMS), while the less massive one (red dot) appears to just evolve off the ZAMS. This evolutionary stage aligns with that observed in other contact binaries, where typically the more massive component exhibits a more advanced evolutionary state. \citep{ChoiZAMS_2016ApJ...823..102C, DotterZAMS_2016ApJS..222....8D}.

\subsection{The O'Connell effect variations}\label{section5.2}
\begin{figure*}[htb!]
	\centering
	\includegraphics[scale=0.5]{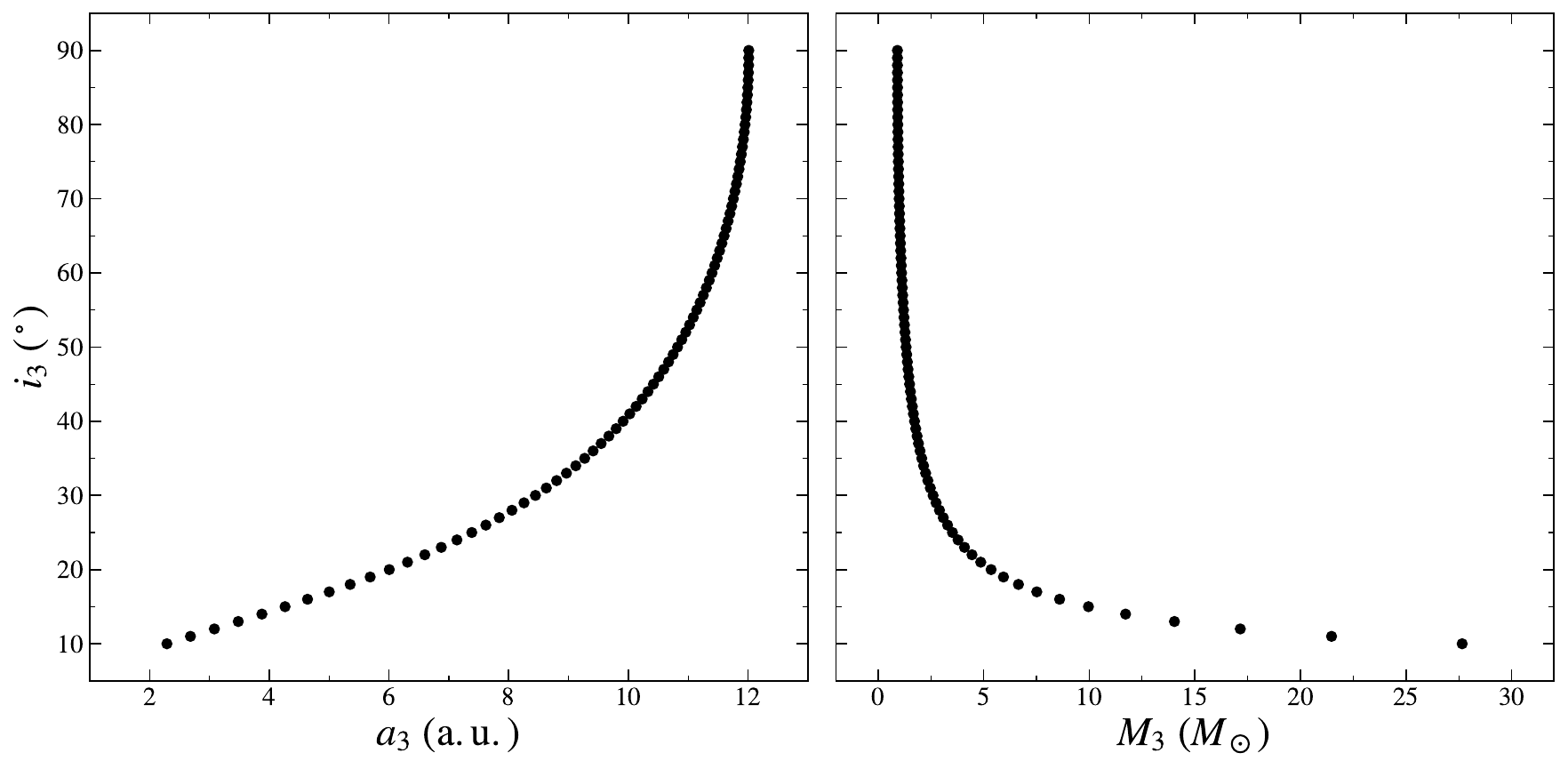}
	\caption{Relation between the orbital radius, mass, and the orbital inclination for an assumed third body in CW Cas.
	\label{figure8}}
\end{figure*}

The O'Connell effect can be divided into two stages, the active stage (high spot activity) and the inactive stage (low spot activity), as noted by \citep{Qian_2014ApJS..212....4Q}. Based on the analysis of photometric time series from TESS light curves (right panel of Figure \ref{figure1}) and the times of maxima shown in panel (a) of Figure \ref{figure4}, we found evidence of a variable O'Connell effect, which could be explained well by our spot model. Similar variations in light-curve maxima have been reported in other contact binaries, such as EPIC 211957146 \citep{Sriram_2017AJ....153..231S}; AD Cnc \citep{Qian_2007ApJ...671..811Q}; CSTAR038663 \citep{Qian_2014ApJS..212....4Q}; OO Aql \citep{LiHua_2016RAA....16....2L}; and V1005 Her \citep{Zhu_2019MNRAS.489.2677Z}, and are commonly attributed to spot activity. To investigate the variations of the O’Connell effect in more detail, we compiled maximum light values (Max I and Max II) from the literature \citep{Jiang_2010PASJ...62..457J, Wang_2014AJ....148...95W}, along with new maxima derived from our observations at Maidanak, YNOs, and from TESS data. Analysis of the combined dataset reveals that the differences between Max I and Max II may exhibit cyclical variations with a period of 349 days, which can be seen in Figure \ref{figure8}. This period variation indicates that the spots activity in CW Cas has a period around 350 days, which is comparable to the spot activity cycle of 466 days reported for the contact binary YZ Phe \citep{Sarotsakulchai_2019PASJ...71...81S}.

\subsection{Mass transfer}\label{section5.3}
Through \mbox{\it O\,--\,C} diagram analysis, we detected a downward parabolic trend superposed on a periodic oscillation. This long-term parabolic term is illustrated by the red dashed line in the (\mbox{\it O\,--\,C}) versus 
$E$ plot of Figure \ref{figure2}, indicating that the orbital period of CW Cas is gradually decreasing. Based on the results of the calculation using equation (\ref{equation2}), we determined that its period decreases at a rate of $\beta=-1.12(8)\times10^{-11}$ day cycle$^{-1}$ and listed it in Table \ref{table3}. This orbital period decrease could be explained by the mass transfer from the more massive to less mass component \citep{Hoffman_2006AJ....132.2260H}. Under the assumption of conservation, its mass transfer rate can be calculated using the following equation \citep{Singh_1986ApSS.124..389S, Zhu_2006MNRAS.367..423Z}:  

\begin{equation}
	\begin{split}
		\frac{\Delta P}{P} = 3\Biggl(\frac{M_{1}}{M_{2}} - 1\Biggl)\frac{\Delta M_{1}}{M_{1}}
	\end{split}
	\label{equation6}
\end{equation}

Based on the derived absolute parameters of CW Cas, we estimate a mass transfer rate of $dM/dt=1.51\times10^{-8}M_{\odot}$ year$^{-1}$.  A comparable mass transfer rate has been observed in many EW-type eclipsing binaries, such as HH Uma \citep{Han_2014NewA...31...26H}, V409 Hya \citep{Na_2014NewA...30..105N} and FZ Ori \citep{Prasad_2014ApSS.353..575P}.

\subsection{Unseen third body around CW Cas}\label{section5.4}

%The appropriate explanation is the presence of a third body. From there search on the orbital period changes of 278 eclipsing binaries, the most plausible explanation of the cyclic period changes is the LTTE through a third body by \citep{LiaoQian_2021MNRAS.508.6111L}. \citet{Sato_2003ApJ...597L.157S} discovered that there are more than 100 planetary companion candidates around G-type stars. \citet{Pribulla_2006AJ....132..769P} suggested that most overcontact binary stars exist in multiple systems. The LTTE through the presence of a third body is a suitable explanation for the cyclic period changes in some overcontact binaries, i.e., V417 Aql \citep{Qian_2003AA...400..649Q}, BI Vul \citep{Qian_2013ApJS..209...13Q}, EQ Tau \citep{LiQian_2014AJ....147...98L} and AH Tau \citep{Yang_2010AJ....139..195Y}. 

Thanks to the open-source data from long-term observations (i.e. DASCH, AAVSO, ASAS-SN etc.), we have extended our \mbox{\it O\,--\,C} curve by over 40\%, and now the data span exceeding 120 years, which allows for tighter constraints on the parameters of the third body. Our \mbox{\it O\,--\,C}  investigation indicates a cyclic variation with a period of $P_{3}=99.4(6)$ yr and a semi-amplitude of $A=0.0628(5)$ d. The data spanning more than one orbital period enhances the reliability of our results compared to previous studies. Taking into account the masses from Table \ref{table6}, We derived the mass function of this system $f(m)=0.131(3)M_{\odot}$ by using the following formula.

\begin{equation}
	\begin{split} 
		f(m) = \frac{(m_{3}\sin i_{3})^{3}}{(m_{1}+m_{2}+m_{3})^{2}} = \frac{4\pi^{2}}{GP^{2}_{3}} \times (a_{12}\sin i_{3})^{3}
	\end{split} 
	\label{equation8}
\end{equation}

\begin{figure*}[htb!]
	\centering
	\includegraphics[scale=0.6]{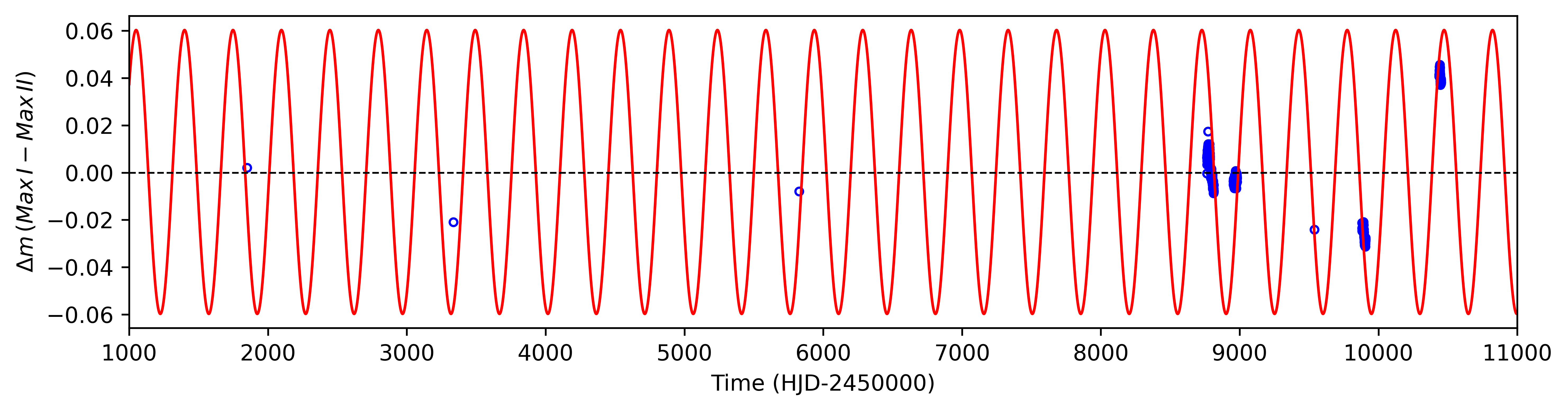}
	\caption{The figure shows a curve fitting of Max I - Max II for the O'Connell effect from, \citep{Jiang_2010PASJ...62..457J}, \citep{Wang_2014AJ....148...95W}, our observation at MAO and TESS data. It is shown that the variations of difference (Max I minus Max II) can be fitted by sinusoidal curve with a period of 349 days and semi-amplitude of 0.033 mag. The dashed line with 0.00 mag means that Max I and Max II are equal without the O'Connell effect.
		\label{figure7}}
\end{figure*}

Assuming the orbital inclination $i_{3}=90^{\circ}$, we estimate the minimum mass of the third body $M_{3}=0.91(1)M_{\odot}$ with the maximal semi-major axis $a_{3}=12.0(2) a.u.$. The relations between orbital inclination, the semi-major axis, and the mass for this third body in CW Cas are shown in Figure \ref{figure7}.

%Assuming that the mass of the third body is close to the masses of components of the contact binary system, it should reveal itself as a third light \citep{Kovalev_2025MNRAS.539.3830K}.
The mass of the third body, approximately equal to one solar mass, is comparable to that of the massive component in the CW Cas system. If it is a main-sequence star, it should be visible in the spectrum and contribute significantly luminosity to total light of the system, like T-Tau0-03027 \citep{Kovalev_2025MNRAS.539.3830K}, V384 Ser and V1038 Her \citep{LiuNP_2023AJ....165..259L}. However, our light curve modeling indicated that the third light contribution ($l_{3}$) is negligible in all bands ($B, V, R_{c},I_{c}$, $TESS$) and spectral observations show no signs of a third component. This unseen third body is most likely a compact object. Consequently, CW Cas resembles other triple systems consisting of a close binary and a compact companion, such as V458 Mon. \citep{Zubairi_2024NewA..11102250Z}, V593 Cen \citep{ZhaoErgang_2019ApJ...871L..10E}, V410 Aur \citep{LiaoWP_2022ApJ...927..183L}, CRTS J154915.7+375506 \citep{WuJinFeng_2024MNRAS.529.3113W}, RR Dra \citep{WangZhiHua_2021MNRAS.507.2804W}). %Similar cases exist in other close binary systems, such as V458 Mon \citep{Zubairi_2024NewA..11102250Z}, where \mbox{\it O\,--\,C} analysis has revealed a massive third component, that was not detected in any photometric observations.}

%only the I band has $L_{3}=0.00151L_{\odot}$. Probably a third-body light reveal in the I band it is a cool dwarf star. However, the light curve is strongly influenced by the magnetic activity of the third light. Since, by mass, it is a main-sequence star, its luminosity must be approximately $0.01L_{\odot}$ \citep{Pecaut_2013ApJS..208....9P}, which corresponds to 1\% of the total luminosity of the system. 
%In this case, the minimum mass of the third component orbiting CW Cas is quite large.

% These two cases help us to conclude that the third body orbiting the CW Cas is most likely an invisible compact object.

%and its semi-major axis $a_{3}=12 a.u$, we did not detect any significant spectral lines of the third body in the spectral profile of the CW Cas. In practice, there are close binary stars, such as T-Tau0-03027 \citep{Kovalev_2025MNRAS.539.3830K}, that exhibit spectral lines from a third body, making it possible to determine the radial velocity of this third component. 

\subsection{Hierarchical quadruple system}\label{section5.5}

Besides the third body, a visual companion GDR3 523846238991809920 (hereafter Vis) near CW Cas was found through analysis of the Gaia DR2 and DR3 data \citep{Gaia_2018yCat.1345....0G, Gaia_2022yCat.1355....0G}. Table \ref{table8}  provides the parallaxes of the CW Cas and Vis from the Gaia DR3 database, which are essentially the same. Moreover, their proper motions are also very close to each other in both the RA. and Dec. directions. Thus, all astrometric parameters indicate that Vis and CW Cas form a gravitational bound system, similar to KM UMa \citep{Meng_2024ApJ...974..103M}. Furthermore, the system radial velocity of CW Cas derived from its RV curve fitting agrees well with that of Vis from Gaia DR3. Therefore we confirm result of \citet{Hartman_2020ApJS..247...66H} who suggested that it is a very wide binary system with Bayesian probability 99\% using Gaia parallaxes and proper motions.  The projected physical separation is around 455.5 a.u., which is much more wider than the unseen third body detected by the LTTE effect. Thus, Vis is the fourth star in this system. Spectral Energy Distribution (SED\footnote{\url{https://github.com/mkounkel/SEDFit} \citep{Kounkel2023}}) analysis of Vis using public photometry suggests that this star is a cool red dwarf, possibly on the main sequence.
%Based on the angular distance  $2.73\mathrm{''}$ acrsec and parallax, it is derived that Vis is about 455.5 a.u. from the binary pair \citep{Hartman_2020ApJS..247...66H}.

\begin{table*}[htb!]
\footnotesize
    \centering
	\caption{Comparison of main parameters and photometric solutions for CW Cas} 
    %\resizebox{\textwidth}{!}{%
	\begin{tabular*}{\textwidth}{@{\extracolsep{\fill}}lccccc}
		\hline\hline
		  Parameters & This paper & \cite{Wang_2014AJ....148...95W} &  \cite{Jiang_2010PASJ...62..457J} & \cite{Pribulla_2001CoSka..31...26P} & \cite{Barone_1988AA...197..347B} \\
		\hline
		$q$ & $1.88(9)$ & $2.057(5)$ & $2.234(2)$ & $0.533$ & $1.84(8)$ \\
		$i$ (degree) & $76.79(27)$ & $81.15(62)$ & $75.91(11)$ & $74.59(11)$ & $73.4(2)$ \\
		$T_{1}$ (K) & $5280$ & $5309$ & $5309$ & $5086$ & $5440$ \\
		$T_{2}$ (K) & $5003(14)$ & $4950(7)$ & $4982(6)$ & $5510(8)$ & $5096(20)$ \\
		%$\mathbf{\Omega_{1}=\Omega_{2}}$ & $\mathbf{5.0055(7)}$ & $\mathbf{5.1990(94)}$ & $\mathbf{5.5400(44)}$ & $\mathbf{2.9317(31)}$ & $\mathbf{4.964(9)}$ \\
		$f (\%)$ & $15.0(12)$ & $22.1(1.6)$ & $6.5(7)$ & $2.2$ & $10.6$ \\
		$M_{1} (M_{\odot})$ & $0.52(4)$ & $0.59$ & $0.56$ & $-$ & $0.58$ \\
		$M_{2} (M_{\odot})$ & $0.98(6)$ & $1.22$ & $1.25$ & $-$ & $1.06$ \\
		$R_{1} (R_{\odot})$ & $0.75(1)$ & $-$ & $0.40$ & $-$ & $0.76$ \\
		$R_{2} (R_{\odot})$ & $1.0(2)$ & $-$ & $0.65$ & $-$ & $1.01$ \\
		$L_{1} (L_{\odot})$ & $0.39(1)$ & $-$ & $0.40$ & $-$ & $0.45$ \\
		$L_{2} (L_{\odot})$ & $0.56(2)$ & $-$ & $0.65$ & $-$ & $0.62$ \\
        $L_{3\it{V}} (L_{\odot})$ & $0.0$ & $0.0942(91)$ & $0.0$ & $-$ & $-$ \\
		$a_{12}$ (a.u.) & $2.25(5)$ & $-$ & $3.00(45)$ & $-$ & $-$ \\
		$K_{1}$ (km s$^{-1}$) & $227.4(5.1)$ & $-$ & $-$ & $-$ & $-$ \\
		$K_{2}$ (km s$^{-1}$) & $120.6(5.7)$ & $-$ & $-$ & $-$ & $-$ \\
		$V_{\gamma}$ (km s$^{-1}$) & $-38.95(3.48)$ & $-$ & $-$ & $-$ & $-$ \\
        Subtype (A, W) & W & W & W & W & W \\
        RV data & yes & $-$ & $-$ & $-$ & $-$ \\
		$P_{3}$ (year) & $99.4(6)$ & $69.9(3.2)$ & $63.7(2.4)$ & $-$ & $-$ \\
		$M_{3\rm{min}} (M_{\odot})$ & $0.91(1)$ & $0.5708$ & $0.31(5)$ & $-$ & $-$ \\
		$a_{3\rm{max}}$ (a.u.) & $12.0(2)$ & $17.44$ & $17.5(3.7)$ & $-$ & $-$ \\
		\hline 
	\end{tabular*}
    %}
	\label{table7}
    %\begin{tablenotes}
			%\item Note: \citet{Pribulla_2001CoSka..31...26P} on his paper used 0.533 
	%\end{tablenotes}
\end{table*}

\begin{table*}[htb!]
	\centering
	\caption{Astronometric parameters and RV for CW Cas and Vis}
	\begin{tabular*}{\textwidth}{@{\extracolsep{\fill}}lcccccccc}
		\hline\hline
		Name & Gaia DR3 & RA & Dec & G$_{mag}$ & Paralax & pm.ra & pm.dec & $V_{\gamma}$ (RV) \\
		       & ID & hh:mm:ss & dd:mm:ss & mag & mas & mas yr$^{-1}$ & mas yr$^{-1}$ & km s$^{-1}$ \\
		\hline
		CW Cas & 523846239000492928 & 00:45:52.70 & +63:05:08.45 & 11.16 & 6.01(1) & -83.56(1)  & 3.05(1) & -38.95(3.48) \\
		Vis & 523846238991809920 & 00:45:52.58 & +63:05:05.84 & 14.91 & 5.68(6) & -82.46(6)  & 1.89(6) & -38.22(3.52) \\
		\hline
	\end{tabular*}
	\label{table8}
\end{table*}

 \par
 Combining all these results, we conclude that CW Cas system is a 2+1+1 hierarchical quadruple system, which comprising a shallow contact binary, a compact object, and a red dwarf. If the compact object originated from a supernova explosion, the continued presence of the contact binary suggests that it remained dynamically stable and was not disrupted or ejected during the event. Thus, the properties of CW Cas can provide valuable constraints on models of compact object evolution \citep{Tauris_1998AA...330.1047T}. Furthermore, the existence of a compact object within such a hierarchical quadruple system renders CW Cas an excellent astrophysical laboratory for investigating the formation and evolutionary pathways of multiple stellar systems.

\begin{acknowledgments}
This work was supported by the International Cooperation Projects of the National Key R \& D Program (No.2022YFF0127300), the Innovation Development Agency under the Ministry of Higher Education, Science and Innovation of the Republic of Uzbekistan, Project (AL-5921122128), the International Partnership Program of the Chinese Academy of Sciences (No. 020GJHZ2023030GC), the Yunnan Fundamental Research Projects (grant No. 202401AS070046, 202503AP140013, 202501AS070055, 2401AW070004) and the Yunnan Revitalization Talent Support Program. We express our gratitude to the staff of the TNO 2.4 m telescope and the Xinglong 2.16 m telescope for their support. We also extend our thanks to the staff of the Sino-Thai 70 cm telescope in Lijiang and 1 m telescopes at Yunnan Observatories for providing new CCD photometric data and assistance of the Maidanak Astronomical Observatory for obtaining multiband photometric data. This work also utilizes photometric and astrometric data from the TESS, ASAS-SN, Gaia, and \mbox{\it O\,--\,C} Gateway light minima times database. The authors express their gratitude to these teams for providing public data. The authors acknowledge the development groups of the SIMBAD database operated at CDS, the NASA ADS Bibliographic Services, and the Mikulski Archive for Space Telescopes (MAST) databases. We used the TESS Input Catalog and Candidate Target List available through the Mikulski Archive for Space Telescopes (MAST) \citep{Tess_input_catalog}, \citep{Stassun_2019AJ....158..138S}. 
\end{acknowledgments}

\begin{contribution}
%%This section gives authors the space to recognize author contributions. The text inside this environment is NOT counted towards the total word quanta. At a minimum, manuscripts are expected to include this text:

%All authors contributed equally to the Terra Mater collaboration.

%% But authors are expected to provide more specific details, e.g. 
%%
%%SC was responsible for writing and submitting the manuscript.
%%WWM came up with the initial research concept and edited the manuscript.
%%OTS obtained the funding and edited the manuscript.
%%EBF provided the formal analysis and validation. He also edited the manuscript.
%%GEH Supervised the undergraduates, wrote the software and administers the project github and Zenodo repositories.
%%
%% Authors can use the Contributor Role Taxonomy (CRediT) at
%% for ideas on how write a good statement tailored to their needs.

\end{contribution}

%\facilities{HST(STIS), Swift(XRT and UVOT), AAVSO, CTIO:1.3m, CTIO:1.5m, CXO}

\software{astropy \citep{Astropy_2013AA...558A..33A,Astropy_2018AJ....156..123A,Astropy_2022ApJ...935..167A},
          LightKurve \citep{Lightkurve_2018ascl.soft12013L}, PyAstronomy \citep{PyAstronomy}, IRAF \citep{IRAF_1986SPIE..627..733T}, TOPCAT \citep{TOPCAT_2005ASPC..347...29T}, Period04 \citep{Period04_2005CoAst.146...53L}, \href{https://www.originlab.com/}{OriginPro}, Version 2024b. OriginLab Corporation, Northampton, MA, USA., Spectral Energy Distribution (SED) fitting \citep{Kounkel2023}, \href{https://ascl.net/2004.005}{PyWD2015} \citep{GuzelPYWD2015_2020CoSka..50..535G}.}   

\bibliographystyle{aasjournalv7}



\end{document}